%% file: ms.tex
\documentclass[iop]{emulateapj}
\pdfoutput=1

\usepackage{graphicx}
\usepackage[usenames]{color}
\usepackage{natbib}
\usepackage{apjfonts}
\usepackage{rotating}

\newcommand{\kms}       {km~s$^{-1}$}

\newcommand{\etal}      {{et~al.}}

\newcommand{\lya}       {Ly$\alpha$}

\newcommand{\push}[1]   {\multicolumn{1}{c}{#1}}
\newcommand{\zqso}      {$z_{\rm{QSO}}$}
\newcommand{\ccz}       {$z_{\rm{CC2}}$}
\newcommand{\ddz}       {$z_{\rm{D1}}$}
\newcommand{\msol}      {$M_\odot$}

\shorttitle{Mg~II Absorption from LRGs}
\shortauthors{Bowen and Chelouche}

\received{April 11, 2010}
\accepted{November 16, 2010}

\begin{document}

\title{The Mg~II cross-section of Luminous Red Galaxies}

\author{David V.~Bowen\altaffilmark{1} and 
Doron Chelouche\altaffilmark{2}
}

\altaffiltext{1}{Princeton University Observatory, Ivy Lane, Princeton, NJ 08544.} 
\altaffiltext{2}{Department of Physics, University of Haifa,
Mount Carmel, Haifa 31905, Israel.}

\begin{abstract}
  
  We describe a search for Mg~II~$\lambda\lambda 2796, 2803$
  absorption lines in {\it Sloan Digital Sky Survey} (SDSS) spectra of
  QSOs whose lines of sight pass within impact parameters $\rho \sim
  200$~kpc of galaxies with photometric redshifts of $z=0.46-0.6$ and
  errors $\Delta z \sim 0.05$. The galaxies selected have the same
  colors and luminosities as the Luminous Red Galaxy (LRG) population
  previously selected from the SDSS. A search for Mg~II lines within a
  redshift interval of $\pm 0.1$ of a galaxy's photometric redshift
  shows that absorption by these galaxies is rare: the covering
  fraction is $f(\rho) \simeq 10-15$\% between $\rho = 20-100$~kpc, for
  Mg~II lines with rest equivalent widths of $W_r \geq
  0.6$~\AA , falling to zero at larger $\rho$. 
  There is no evidence that $W_r$ correlates with impact
  parameter or galaxy luminosity.  
  Our results are consistent with existing scenarios in which 
  cool Mg~II-absorbing clouds may be absent near LRGs because of the
  environment of the galaxies: 
  if LRGs reside in high-mass groups and clusters, either their
  halos are too hot to retain or accrete cool gas, or the galaxies
  themselves --- which have passively evolving old stellar populations ---
  do not produce the rates of star formation and outflows of gas
  necessary to fill their halos with Mg~II absorbing clouds.  In the
  rarer cases where Mg~II is detected, however, the origin of the
  absorption is less clear. Absorption may arise from the little cool
  gas able to reach into cluster halos from the intergalactic medium,
  or from the few star-forming and/or AGN-like LRGs that are known to
  exist.

\end{abstract}


\section{Introduction}

Red galaxies contain the bulk of all the stellar mass that exists
in the local universe \citep{hogg02}, and thus mark the conclusion of
both the dominant star formation processes, and 
of the assembly of galaxies.
The Luminous Red Galaxies (LRGs)
are a population selected by their optical colors to contain old, passively
evolving stellar populations at intermediate redshift. The first two
principal samples of LRGs were constructed from the {\it Sloan Digital Sky
Survey} \citep[SDSS;][]{york00} for galaxies at redshifts $z\approx 0.2-0.4$
\citep{eisenstein01}, and from the {\it 2dF-SDSS LRG And QSO} (2SLAQ)
survey \citep{cannon06} for galaxies at $z\approx
0.4-0.7$. Subsequently, the
properties of the LRGs have been investigated by many authors: LRGs have
luminosities $L > L^*$ and large stellar masses $M \ga 10^{11}
M_\odot$ \citep{banerji10};  the majority of LRGs show spectral
energy distributions (SEDs) with little evidence for on-going star
formation 
\citep{roseboom06}; they
reside in dark matter (DM) halos with masses $M \simeq 10^{13-14}
M_\odot$ \citep{mandelbaum06}, and in Halo
Occupation Distribution (HOD) models, most LRGs are central galaxies
dominating the DM halos \citep[e.g.][and refs therein]{zheng09}; they
appear to be
most prevalent in clusters \citep{ho09}.
At $z\sim 0.2-0.8$, the evolution of the LRG luminosity function (LF)
is found to be consistent with a mode in which stars form at high-$z$,
and evolve passively thereafter with little further star formation
\citep[e.g.][]{wake06,brown07,cool08,banerji10}.  It appears that these
massive galaxies have already been assembled by $z\sim 1$
\citep{conselice07,ferreras09, banerji10}.

Given the colloquial view of LRGs as ``red and dead'', they would seem
an unlikely population of galaxies to harbor significant amounts of
cool gas. Nevertheless, there has recently been considerable interest
in using the clustering of LRGs to infer the masses of halos in which
Mg~II absorption lines arise \citep[][--- see
  \S\ref{sect_why_noabs} for a detailed discussion of these
  results]{bouche06, tinker08, gauthier09,lundgren09}. 
The Mg~II~$\lambda\lambda 2796,2803$ doublet is the first strong UV
transition to be redshifted into the optical wavelength regime, at $z
\ga 0.2$ --- a redshift low enough for galaxies responsible for the
absorption to be identified. Historically, initial searches for galaxies at the same
redshifts as previously identified Mg~II systems were quick to confirm
an association
\citep{bergeron86,yanny87,crist87,berg_boiss91,lebrun93}. More
extensive surveys suggested that all galaxies with luminosities $\sim
L^*$ have absorbing cross-sections of radii
$R_g\:\sim\:60$~kpc\footnote{Throughout 
this paper, we adopt
$H_0=70$~\kms~Mpc$^{-1}$, $\Omega_m\:=\:0.3$ and
$\Omega_\Lambda\:=\:0.7$. 
Distances cited from published works
that are originally given in units of $h^{-1}$~kpc 
are converted to this cosmology.}
 with
unity covering fraction $f_c$
\citep[e.g.][]{steidel93conf,stei95conf,steidel_mg2summ}.
The subsequent development in numerical hydrodynamical simulations,
and the refinement of semi-analytical modelling, has provided a
natural explanation for multi-phase gas in galactic halos, through the
inflow of baryons into regions of high density. Particularly appealing
for the origin of Mg~II systems are those models which describe cold
flows along DM filaments, where unshocked gas at temperatures of $<
10^5$~K are channeled into galaxies \citep[e.g.][and refs
therein]{keres09,brooks09,kaufmann09} at the same time that shocked
gas is accreted by the halo. As Mg~II absorption is thought to arise
in photoionized gas with temperatures $\sim 10^4$~K, the existence of
halos with Mg~II absorbing clouds is consistent with these
simulations. However, one of the more important parameters
  governing the existence of cool gas in DM halo models is the mass of the
  halo. The simulations predict that in halos with masses $>10^{13-14}
  M_\odot$, where the LRGs reside, there should be no cool gas,
  whereas in lower mass halos, where the $L^*$
  galaxies mentioned above are found, there could still be enough cool
  gas to cause Mg~II absorption along sightlines that intercept the
  halos (see \S\ref{sect_why_noabs}). In these scenarios, QSO
  sightlines passing close to LRGs should show no Mg~II absorption.

On the other hand, recent studies have focused less on the
  relationship of Mg~II absorption and halo mass,
and more on the idea that Mg~II
systems originate in some combination of galactic disks and starburst
driven outflows. Galactic outflows 
offer a possible (although observationally unsubstantiated)
explanation for finding cool gas clouds out
to many tens of kpc from a galaxy. 
The fact that large-scale outflows are driven by significant star
formation activity has led to more detailed investigations into the
link between Mg~II systems and late-type galaxies.  \citet{bouche07},
for example, showed that two-thirds of 21 sampled Mg~II systems with
rest-frame equivalent widths (REWs) of $W_r > 2$~\AA\ have H$\alpha$
emission from galaxies with star formation rates of $1-20
M_\odot$~yr$^{-1}$.  \citet{zibetti07} found excess (broad-band) emission
around stacked SDSS QSO images with known Mg~II systems,
compared to stacked images of QSOs without absorption. The
optical colors of the light suggested that the SEDs of strong Mg~II
absorbers are bluer than for weak ones, indicative perhaps of the
domination by blue star-forming galaxies to the production of strong
lines.  Moreover, the large sky coverage of the SDSS, combined with
the remarkable number of QSO spectra obtained, has made it possible
to examine the relationship between star forming galaxies and Mg~II
systems, without regard to the identification of any galaxy close to
the line of sight. Emission lines have been detected in QSO spectra at
the same redshifts as absorption line systems, either in stacks of
spectra \citep{wild07,menard10} or in rarer cases, individual QSO
spectra \citep{noterdaeme10,sanch10}.

Whether these results indicate that absorption arises in large-scale
gaseous outflows, rather than simply from galactic disks that host
star forming regions, is less clear.  The idea that galactic disks are
responsible for Mg~II lines has long been appealing \citep{wagoner67,
  bowen_95, charlton96, steidel_disks}, and modeling Mg~II absorbing
gas as a combination of disk and outflow features, naturally explains
the range of galaxy luminosities and star formation rates observed
\citep{chelouche10}.  The detection of low ionization optical
absorption lines in outflows from nearby starburst galaxies, using the
light from the galaxy itself to probe the gas
\citep{martin99,schwartz04,martin05,rupke05} adds support for the
model, and indeed, Mg~II absorption is seen from outflows of high-$z$
galaxies \citep{tremonti07,weiner09}. Detailed investigations of
starburst galaxies certainly suggest that these types of galaxies
produce multi-phase outflows \citep{strickland04,strickland04b} that
would show absorption lines from many different ionization states if
penetrated by a QSO sightline.  It remains to be seen though how
ubiquitous these outflows are at higher redshift around the majority
of late-type galaxies, whether QSOs lines of sight, probing in a
direction transverse to galaxy sightlines, would see similar
absorption signatures as those recorded toward the centers of
galaxies, or, indeed, whether outflowing gas capable of producing the
types of absorption seen can escape out to many tens of kpc from galaxies.

The redshifts at which Mg~II becomes detectable in SDSS spectra [and
where the SDSS sensitivity allows for detection of weaker lines in
higher signal-to-noise (S/N) data] are in the range $z \simeq
0.36-0.43$ (3800$-$4000~\AA ).  An interest in the association between
Mg~II systems and LRGs arises from the fact that at these redshifts
and beyond, the LRGs are, pragmatically, the only galaxies detectable
in SDSS images with reliable photometric redshifts 
(hereafter ``photo-$z$s'').  The availability of good quality photo-$z$s
can be used to cull
a large number of faint,
intermediate-$z$ galaxies without having to first measure their
redshifts through expensive, time-consuming spectroscopic surveys.
If the galaxies are observed as part of a survey --- such as SDSS ---
which also records the spectra of a similarly large number of QSOs,
then the opportunity to study the absorbing properties of the galaxy
ensemble is obvious.   
The fact that, in this particular case, such a set of galaxies happens
to be
LRGs, means we can explore Mg~II absorbing gas in the halos of a restricted,
specific, and rare type of galaxy and their environment.
The theoretical considerations
discussed above suggest that we should find no absorption, but this paper
seeks to test that prediction.

Most surveys designed to find galaxies responsible for absorption
systems have been based on targeting Mg~II systems known {\it
  ab initio}. While this approach yields information on the origin of
the absorption, it does not necessarily define the absorbing
properties of galaxies.  Prior to the launch of the {\it Hubble Space
  Telescope} (HST), efforts to study the gas in disks and halos of
galaxies focused on pairing very low-redshift galaxies with QSO
sightlines and searching for absorption lines (Ca~II~$\lambda\lambda
3933, 3968$, Na~I~$\lambda\lambda 5889,5895$, etc.) that could be
observed from the ground
\citep[e.g.][]{boksenberg78,boksenberg80,blades81,bergeron87,
  womble90,BPPB91}. With HST, it became possible to search for the
more sensitive and abundant UV lines from nearby galaxies
instead. That work showed that at $z\sim 0$, there is
little evidence for Mg~II absorbing clouds around galaxies beyond
$\sim 50$~kpc \citep{bowen_95}, and when detected, the absorption
originates in galactic disks \citep{bahcall92, bowen_95,bowen_m61} or
in tidal debris between the identified galaxy and an often fainter
companion \citep{bowen_93j,bowen_95} --- even when the targeted galaxy
is a known starburst \citep{norman96}.

Unfortunately, studies of galaxies in the local universe are hampered
by the necessity of having to use QSOs with the brightest UV continua,
in order to obtain sufficient S/N with the available HST
spectrographs.  To avoid this limitation, and to make use of the huge
number of QSO sightlines observed spectroscopically with SDSS, more
aggressive attempts have been made recently [following a much earlier
study by \citet{bechtold92}] to identify galaxies at higher redshifts
and to again search for Mg~II lines shifted into the optical, without
any prior information on the absorption lines in the background QSO
spectra.  The absorbing cross-section of a galaxy, $R_g$, and the
covering fraction of the gas $f_c$, are obviously inter-related; if
$f_c$ declines with radius (i.e. if there is no sharp boundary to $R_g$),
then $R_g$ can be set arbitrarily for a given
$f_c$, and vice versa. Similarly, both $R_g$ and $f_c$ can, in
principle, depend on the equivalent widths of the lines.  In a study
using early results from SDSS, \citet{tripp_china} found galaxies at
$z\sim 0.3-0.6$ with $f_c$ only $\simeq 50$\% for $R_g \sim 50$~kpc,
for sensitivities of $W_r \ga 0.1$~\AA . Similarly, for $W_r \geq
0.3$~\AA , \citet{barton09} found $f_c \simeq 0.25-0.4$ for $R_g
\simeq 110$~kpc, or $f_c \sim 0.25$ for smaller radii of $R_g \sim
50$~kpc, for galaxies at $z\simeq 0.1$.  Conversely, however,
\citet{chen10} [see also \citet{chen08}] measured much higher covering
fractions and larger cross-sections for galaxies at $z\simeq 0.2$:
$R_g\simeq 110$~kpc and $f_c \simeq $[0.7,0.8] for $W_r \geq$
[0.3,0.1]~\AA .  These radii are obviously considerably larger than the
cross-sections measured for galaxies at $z\sim 0$.

One caveat to this work is that at higher redshifts, identifying the
processes responsible for absorption systems is much harder.  At these
redshifts, galaxy features are hard to discern even in high resolution
HST images, low surface brightness features are difficult to detect,
multi-wavelength observations (e.g. 21~cm
interferometry observations, which are important for mapping high H~I
column densities around galaxies) are often not possible,
and the light of a background QSO masks any faint emission directly
along a sightline.  (Indeed, these difficulties also hamper the study
described in this paper, as we discuss in \S\ref{sect_interlope}.)
Hence, even with more comprehensive surveys of absorbing galaxies at
$z\ga 0.1$, it remains difficult to know whether Mg~II absorption
marks gas inflowing from filaments, outflows from star formation
activity, the extension of galactic disks, the tidal debris between
multiple galaxies, or some combination of all these mechanisms.
Hence 
our current ambitions are to pre-select galaxies with defined
characteristics, whose properties have been determined from understood
processes, and use
such calibrations to interpret the origin of the observed absorption
systems.

In this paper, we study the absorbing properties of LRGs ---
again, selected without any prior bias toward knowing whether or not
Mg~II exists at the redshifts of the galaxies. Our initial approach
was to simply define a sample of SDSS galaxies close to QSO sightlines
that had photometric redshifts with very small errors, and to search
for Mg~II absorption in the SDSS spectra of the background QSOs. As
described below, we began such a program without regard to the type of
galaxies selected. We found, however, that the selection of LRGs
in the SDSS arises naturally from demanding that the
photo-$z$ errors of galaxies be small.  
We note here that, 
concurrent with our
investigations, \citet{gauthier10} have also examined the absorbing
properties of a set of galaxies (with photo-$z$s) selected {\it a
  priori} to be LRGs. Their results, obtained independently of ours,
obviously offers an important comparison with our study, and we
discuss their results in \S\ref{sectResults} and
\S\ref{sect_discussion}.

The outline of this paper is as follows: in \S\ref{sect_sample} we
define how we selected the galaxies to study, and describe the search
for Mg~II at their photometric redshifts in \S\ref{sect_abs}. In
\S\ref{sect_gals} we describe the properties of the galaxies, and
demonstrate why the majority of them are LRGs. We discuss the results
of the study in \S\ref{sectResults}, and their implications in
\S\ref{sect_discussion}.

\section{Sample Selection}

\subsection{Selection of galaxies with photometric redshifts }\label{sect_sample}

We began an investigation of galaxies close to SDSS QSOs with no 
regard to the properties of the galaxies themselves. That is, there
was no initial aim of selecting only LRGs in our study. We
aimed to utilize both the photometric redshift information that was
available for SDSS galaxies, and 
existing high quality spectra of
SDSS QSOs. Only after forming a suitable sample of QSO-galaxy pairs
did it become clear that the galaxies were LRGs. We discuss the
properties of the galaxies in our sample more fully in
\S\ref{sect_gals}. In this section, we describe the criteria that
were actually used to select galaxies close to QSOs.

We first compiled a list from DR6 of QSOs observed spectroscopically
by SDSS that had galaxies within 30 arcsec of the QSO
sightline\footnote{That is,  objects which are identified as ``neighbors''
  to QSOs in the specObj view of the SDSS database} and that also had cataloged
photometric redshifts. We imposed a basic cutoff for the photo-$z$ of
a galaxy to be \ccz$> 0.4$ (see below for a definition of this
quantity), \ccz$< $\zqso , and a galaxy morphology (``type'' $= 3$). The
minimum photo-$z$ of \ccz$=0.4$ was imposed since Mg~II was not
expected to be readily detectable below $\simeq 3900$~\AA\ in SDSS
spectra (see below). This basic selection yielded a sample of $\sim 80,000$ pairs.

The most important criterion in selecting a suitable subset of
galaxies for which we would search for Mg~II absorption, was that the
photo-$z$ should be reliable --- that is, the photo-$z$ error should
be small.  The SDSS database contains two measures of a galaxy's
photo-$z$ estimated from the work of \citet{oyaizu08}, tagged in the
SDSS database as ``D1'' and ``CC2'' (which we refer to as $z_{\rm{D1}}$ and
\ccz ), both neural network based estimators. $z_{\rm{D1}}$ uses galaxy
magnitudes in the photo-$z$ fit, while \ccz\ uses only galaxy
colors. Each of these photo-$z$ estimates have errors associated with
them (based on an empirical ``Nearest Neighbor Error'' method),
although there are two measurements of the error in each case, because
the distribution of errors $\Delta z = |z_{\rm{phot}} -
z_{\rm{spec}}|$ --- the difference between the photo-$z$ and the real
redshift of a galaxy in a training set of galaxies --- are
non-Gaussian. There are two estimates in the error of the photo-$z$s: 
one error is given by $\sigma_{68}$,
which contains 68\% of the values of $\Delta z$ in the training
sample; the other, which we call $\sigma_{\rm{rms}}$ (simply
  ``$\sigma$'' in Oyaizu~\etal ), is the rms scatter (again, see
Oyaizu~\etal\ for more details). 

The accuracies of CC2 and D1 decline quickly for $r-$band galaxy
magnitudes fainter than $r=20.0$. The majority of galaxies found close
to QSOs in our initial search for QSO-galaxy pairs had $r> 20.0$, and
excluding these would result in a sample too small to be of interest.
Based purely on galaxy magnitude, the errors in $z_{\rm{D1}}$ and
\ccz\ rise sharply from 0.05 to 0.09 between $20.0 < r < 21.0$ (see
Fig.~5 of Oyaizu~\etal ).  Fortunately, for the galaxy redshifts we
are interested in --- at $z > 0.4$ --- photo-$z$s are defined to a
high accuracy as the 4000~\AA\ break in the galaxy spectra passes
through the $r$-band.  Figure~6 of Oyaizu~\etal\ shows that between
$z=0.4-0.6$, $\sigma_{68}$ for \ccz\ ranges from $\simeq 0.04-0.09$,
and is smaller than $\sigma_{\rm{rms}}$ by $\Delta z \approx 0.02$.
In this paper, we adopt $\sigma_{68}$ as the error in a galaxy's
photo-$z$, and denote this simply as $\sigma(z)$.

We interpret Oyaizu~\etal 's results to mean that, for 68\% of selected
QSO-galaxy pairs, the true redshift of the galaxy lies between
$z_{\rm{phot}} - \sigma(z)$ and $z_{\rm{phot}} + \sigma(z)$.  From our
original collation of galaxy pairs, we therefore selected only
galaxies with $\sigma(z) \leq 0.05$. We took the maximum photo-$z$
to be 0.6, while the minimum photo-$z$ was not determined by the photometric
properties of the galaxies themselves, but by the minimum wavelength
so that Mg~II could be detected in SDSS spectra. We took this minimum to
be 3800~\AA, plus a redshift path which allowed a search for Mg~II
within $2\sigma(z)$ (see below). The resulting minimum redshift
was 0.46. Given this redshift range, and the 30 arcsec limit for
finding galaxies, the maximum impact parameters studied are between
175 and 201~kpc.

To further exclude galaxies with spurious photo-$z$s, we also
demanded that $|$\ccz$-$\ddz$| < \sigma(z)$, i.e. that the
photo-$z$s estimated from the two different methods agreed to within
$1\sigma$ of each other. Additional constraints included:
\zqso$-$\ccz$ > 0.3$, to avoid Mg~II associated with the QSO itself;
and \zqso$< 2.1$, to exclude high-$z$ \lya -forest lines in the
SDSS spectra. In a few cases, more than one galaxy satisfying all
these criteria was found in a single QSO field; these galaxies were
flagged and are discussed later.

Further culling of the original sample was made based on the
spectroscopic properties of the background QSOs. These are discussed
in the next section.

\subsection{The search for Mg~II absorption from foreground galaxies }\label{sect_abs}

\begin{figure}
\vspace*{-1.75cm}\hspace*{-0.3cm}\includegraphics[width=14cm]{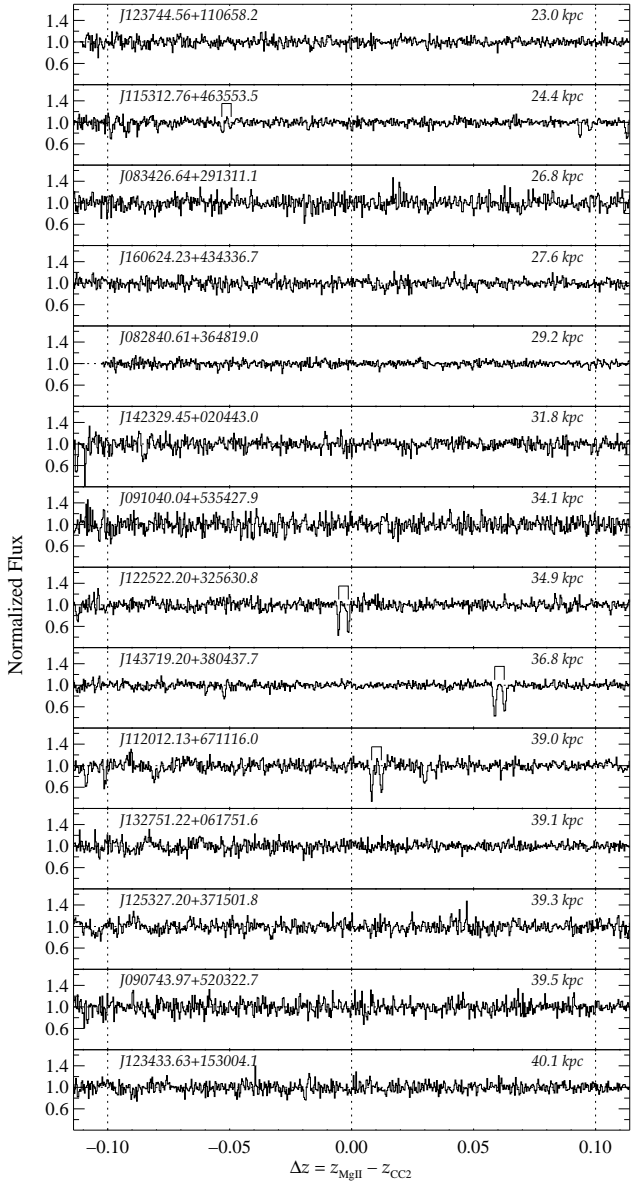}
\caption{\label{fig_spectra} Representative normalized spectra of QSOs in our
  sample with sensitivities $W_{\rm{lim}} < 0.4$~\AA .  Spectra
  are selected for QSOs that have the 14 closest galaxies, with their
  impact parameters $\rho$
  printed at the top right of each panel. 
  The $x$-axis shows the wavelength reduced to a redshift range for
  Mg~II~$\lambda 2796$, and normalized to \ccz .
The QSO designation is given
at the top left of each panel. Mg~II, when detected, is marked with
two vertical bars. The dotted lines at $\Delta z = \pm 0.1$ represent
the range over which Mg~II was searched for. }
\end{figure}

For the QSO-galaxy sample described above, formed from criteria
applied to galaxies, we next considered a set of criteria based on
the suitability of the QSO spectra. We first rejected all QSOs with
S/Ns in the continuum of less than 6.0 per
pixel at \ccz , based on the error arrays generated from the pipeline
extraction of the one-dimensional SDSS spectra. At $z\simeq 0.5$, this
S/N corresponds to a 3$\sigma$ REW limit
of $W_r \approx 0.7$~\AA\  (when calculated using the method
described below). 
We rejected spectra with extraction defects, Broad Absorption Line
complexes, complicated continua, or those
contaminated by complex and numerous higher-$z$ systems. In some
cases, we rejected spectra where Mg~II was expected to fall close to a
particularly sharp QSO emission line, since defining the continuum at the peak of
these emission lines can be difficult (particularly in cases where
associated absorption lines are seen superimposed on the emission
line).

For the accepted spectra, we normalized the QSO continua by defining a
least-squares fit of Legendre polynomials to the continuum flux
\citep{sembach92}. For each spectrum, we then calculated 
  an array of observed $2\sigma(W_i)$ EWs, where,
  at each pixel $i$, $\sigma^2(W_i) = \sum^{n}_i\: (\sigma_i\:\delta
  \lambda )^2$, and $\delta\lambda$ is the wavelength dispersion in
  \AA~pixel$^{-1}$.  Error arrays from the SDSS pipeline again
provided values of $\sigma_i$, and we took $n$ to be a value
twice the resolution expected for SDSS
spectra\footnote{ At the wavelengths where Mg~II is expected the
  resolution of the SDSS spectra is $\simeq 2.2$~\AA\ at 4500~\AA,
  FWHM, or $\sim 150$~\kms .}, or five pixels.

In each normalized QSO spectrum, we searched for Mg~II absorption
within a window of $\pm 2\sigma(z)$ of \ccz\ , i.e. $\Delta z = \pm
0.1$. Using a value of $\pm
  2\sigma(z)$ corresponds conveniently to assuming that $\approx 95$~\% of
  the LRG photo-$z$s will be correct; if we had chosen to use
  $\sigma_{\rm{rms}}$ instead of $\sigma_{68}$, we would have to have
  increased the path length by $\delta z \approx \pm
0.04$ to include $\pm 2\sigma_{\rm{rms}}$. Reviewing the SDSS
QSO spectra used in our survey indicates that one more Mg~II system would have been
detected if we had adopted this estimate of the photometric redshift
errors. As we shall see below, this would have made little difference
to the final results.

Detection of Mg~II required that both members of the doublet be found
with EWs $> 2\sigma(W_i)$,
although only the
Mg~II~$\lambda 2796$ line had to lie in the $\pm 2\sigma(z)$ window.
When detected, we
fitted Gaussian profiles to the lines and calculated a final redshift
for the absorption system by averaging the redshifts measured for each
line of the doublet. Line EWs were measured in the
standard way, $W = \sum^{n}_i\: (1-F_i) \:\delta \lambda$, where $F_i$
is the normalized flux at the i'th pixel, and the sum is calculated
over $n$ pixels. As noted above, for unresolved lines,
  $n=5$. However,
many strong Mg~II lines found in SDSS spectra have widths that
are larger than this; in order not to under-estimate $W$ therefore, for lines
detected with EWs $> 6 \sigma(W_i)$ (i.e. lines defined
with high S/N),
we increased $n$ to cover a width of $\pm 3
\sigma_{\rm{gauss}}$, where $\sigma_{\rm{gauss}}$ was the width of
the Gaussian profile fitted to a line.

Were the redshift of a galaxy to be measured
  spectroscopically, the $2\sigma$ EW limit for Mg~II if a doublet was
  {\it not} detected would be given by the value of $2\sigma(W_i)$ at
  the appropriate wavelength. In order to define an EW
  limit over a wavelength range corresponding to $ \pm 2\sigma(z)$,
  however, we need to be sensitive to lines with EWs greater than some
  EW limit over the entire range of $\pm 2\sigma(z)$. We therefore
  define the $2\sigma$ REW limit, $W_{\rm{lim}}$, to be the maximum
  value of $ 2\sigma(W_i)$ in the EW array between $\pm 2\sigma(z)$,
  after each point in the EW array is corrected to be in the rest
  frame of $z_i = \lambda_i/\lambda_0 -1$ (where $\lambda_0$ is the
  wavelength of the Mg~II~$\lambda 2796$ line).  For SDSS spectra,
  $\sigma(W_i)$ is relatively flat as a function of wavelength over
  the redshift range considered here, so is close to the value that
  would be calculated at exactly \ccz .

\begin{figure}
\vspace*{-1cm}\hspace*{-1.5cm}\includegraphics[width=13cm]{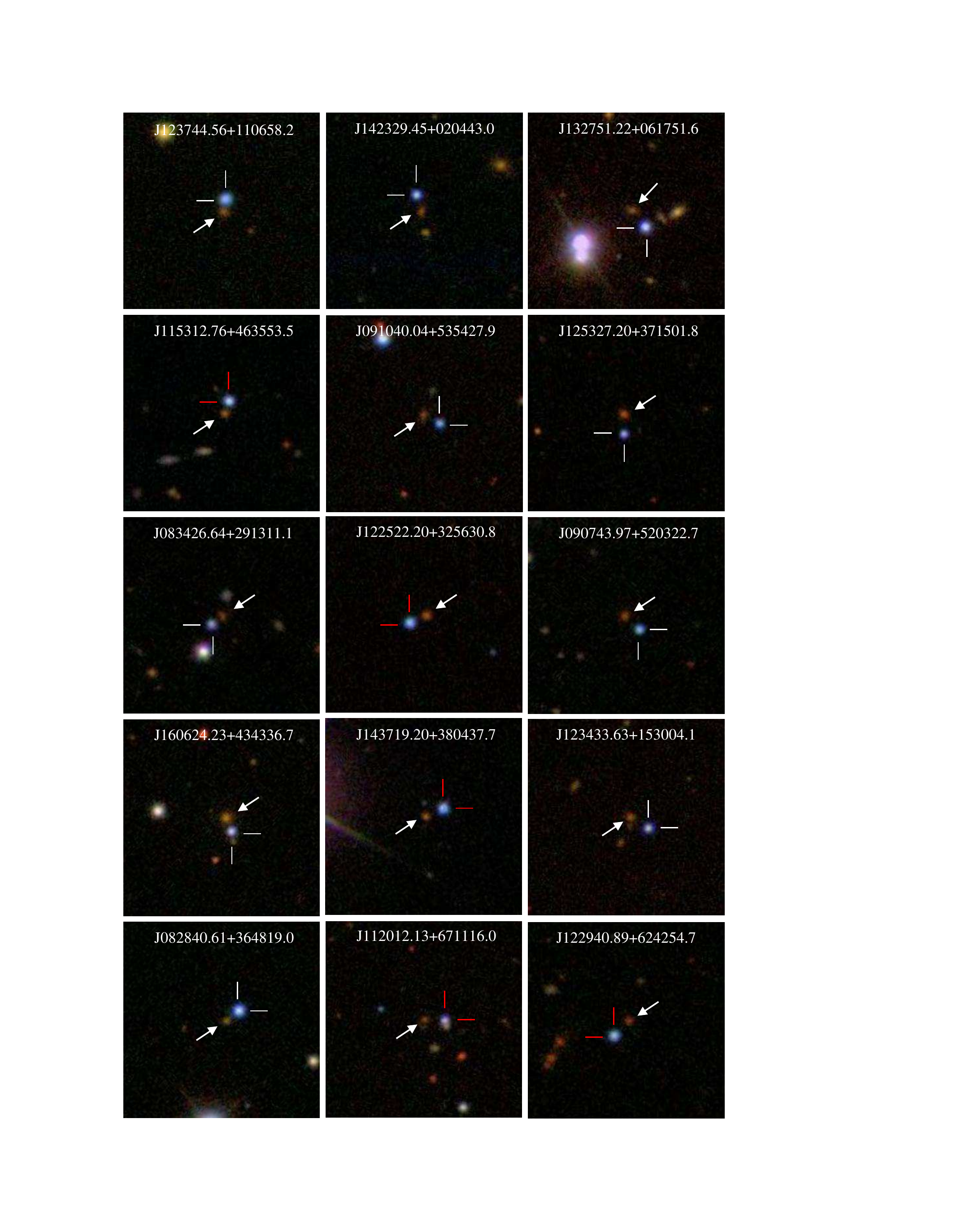}
\vspace*{-1cm}\caption{\label{fig_pairs} Representations of some of the SDSS
  QSO-galaxy pairs 
  used in this survey. The fields shown include
  those with QSO spectra plotted in Figure~\ref{fig_spectra}; they thus represent
  the pairs with the smallest impact parameters, and with
  $W_{\rm{lim}} \leq 0.4$~\AA\ for the QSO spectra. Each image is 60 arcsec on a
  side, with NE to the top-left. Although not labeled, the impact parameters between QSO
  sightlines and galaxies are shown in Figure~1 and are between 20.0 (top left corner image) and 40.0~kpc
  (bottom right corner image). Full details can be found in
  Table~\ref{tab_sample}. The QSO
  designation is printed at the top of each image, and its position
  is indicated by cross-hairs.  Cross-hairs in red represent QSOs that
  show Mg~II near the photo-$z$ of the galaxy, while white cross-hairs
  indicate that no absorption was detected. Galaxies are identified
  by arrows, and always lie at the center of the field.
}
\end{figure}

Examples of the spectra used in our survey are shown in
Figure~\ref{fig_spectra}. For this figure, the QSOs chosen are sorted
by QSO-galaxy impact parameter, and we have only selected spectra with
sensitivities of $W_{\rm{lim}} \leq 0.4$~\AA , simply to
highlight better quality data.  Figure~\ref{fig_pairs} shows color
representations of the fields around these QSOs. Each image is a
cutout extracted from the SDSS database, and so is a $g-$, $r-$, and
$i-$band composite of the original imaging data \citep{lupton04}. Full
details on all the pairs selected are given in Table~\ref{tab_sample};
the table includes details of the QSOs and galaxies in the sample, as
well as the results from the search for Mg~II absorption: values of
REWs when Mg~II was detected, or limits when the lines were not
detected, are listed therein.

\subsection{Galaxy Properties }\label{sect_gals}

\begin{figure}
\vspace*{-1.75cm}\hspace*{-2cm}\includegraphics[width=13.5cm]{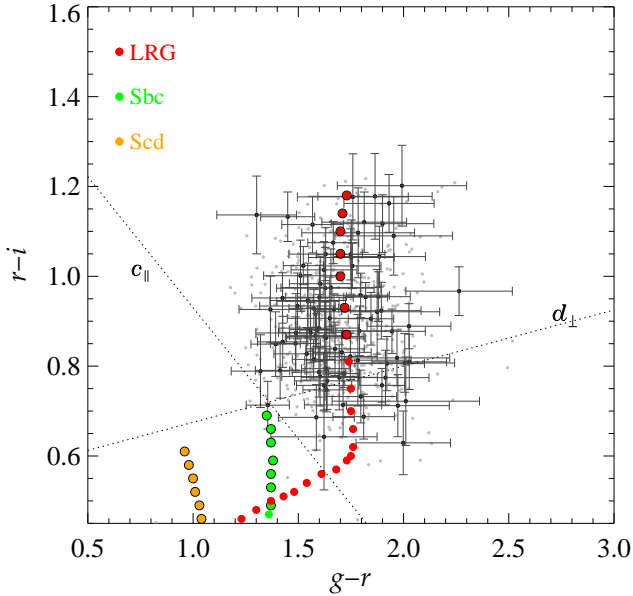}
\caption{\label{fig_colors} Plot of broad-band colors for galaxies in
  our selected sample of QSO-galaxy pairs. Black circles represent
  galaxies for which Mg~II was detected within $\delta z \pm 0.1$ of
  the photo-$z$ of the galaxy, \ccz ; gray circles represent
  non-detections of Mg~II. To avoid confusion, errors in the colors
  are shown only for Mg~II detections, and not for Mg~II
  non-detections.  Also shown are theoretical colors for a Luminous
  Red Galaxy (LRG, in red), an Sbc (green) galaxy, and an Scd (orange)
  galaxy, calculated over redshift intervals of 0.02. The last points
  plotted are for $z=0.6$, and for emphasis, the redshifts
  corresponding to the criterion used to select galaxies,
  \ccz$=0.46-0.6$, also have black outlines around the circles. The
  color cuts used by \citet{cannon06} are shown as dotted lines, and
  are $c_\parallel = 0.7(g-r)+1.2(r-i-0.18)\geq 1.6$, and $d_\perp =
  (r-i)-(g-r)/8.0 > 0.55$ (see text).  }
\end{figure}

In Figure~\ref{fig_colors} we plot the $g-r$ and $r-i$ colors of the
galaxies in our sample. Even though the colors of galaxies were not
used in their selection, their distribution in the color diagram is
confined to a narrow region with $1.5 \le g-r \le 2.0$ and $0.7 \le
r-i \le 1.2$. These colors are similar to those used by
\citet{eisenstein01} and \citet{cannon06} to define
LRGs at the same redshifts as the galaxies in
our sample.
The reason that these authors applied such color cuts initially, and the
reason for expecting their surveys to focus on $z\sim 0.2-0.5$ early
type galaxies, was because the SEDs of such galaxies are predicted to
fall above specific color cuts.  So, for example,
Figure~\ref{fig_colors} shows the cuts, designated
$d_\perp$ and $c_\parallel$, used by Cannon~\etal\ 
to define spectroscopic targets in the 2SLAQ sample. 
They defined $d_\perp$ to select early
type galaxies at increasingly higher redshifts, and $c_\parallel$ to
eliminate late-type galaxies altogether.  In the figure, we have drawn
a line for $d_\perp = 0.55$, which defined ``sample 3'' of the
galaxies studied by Cannon~\etal\ and which contained the bulk of
their LRG sample. Most of the galaxies we selected lie in the same
color region as their LRGs.  

The way in which the SEDs of early-type galaxies
produce different broad-band colors as a function of redshift is well
understood. A change in the $g-r$ color to a constant value at $g-r
\simeq 1.7$ measures the transition of the 4000~\AA\ break in a
galaxy's SED as it moves from the $g$-band to the $r$-band at $z\sim
0.4$; $g-r $ begins to increase again at $z = 0.7$ as the 4000~\AA\
break moves into the $i$-band. To demonstrate this effect, we have
used the LRG SED provided by \cite{blanton07} to show how the colors
change with $z$ when convolved with SDSS filters. These are shown as
red circles in Figure~\ref{fig_colors}, where a circle is plotted
every $\delta z = 0.02$ up to $z=0.6$. We have also plotted the
expected colors of a late-type Sbc and Scd galaxy using the SEDs of
\citet{coleman80}, as green and orange circles, respectively.

Figure~\ref{fig_colors} shows that the colors of the galaxies in our
sample are largely consistent with early-type LRG colors, and not with
late-type galaxies. More sophisticated modeling shows that
theoretical SEDs can also reproduce these colors, assuming stars formed
in a single burst 11~Gyr ($z\sim 2.5$) ago and with a star formation
timescale 
of $1-3$~Gyr \citep[][and refs.~therein]{banerji10}.

Hence, the majority of the galaxies in our sample are early-type LRGs.
The reason for principally selecting LRGs
is almost certainly because these types of galaxies, with a well
defined 4000~\AA\ break, provide the most accurate photo-$z$
estimates. Since we selected galaxies based on small values of
$\sigma(z)$, we therefore predominantly picked galaxies with the strongest
Balmer breaks.

\begin{figure}
\vspace*{-1.5cm}\hspace*{-2.25cm}\includegraphics[width=12cm]{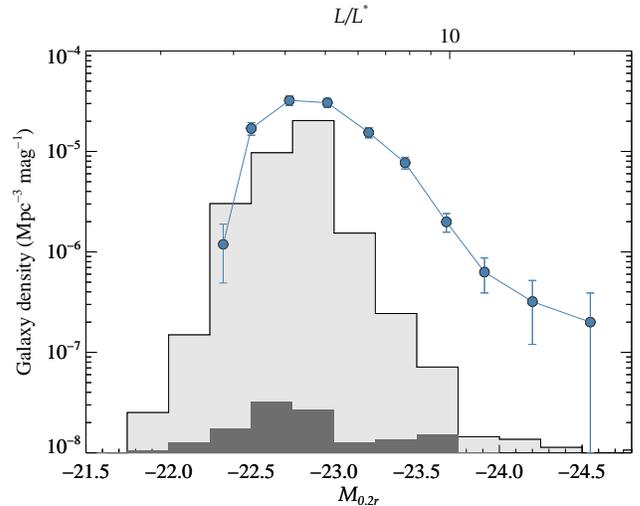}
\caption{ \label{fig_lf}
Plot of the $r$-band luminosity function of
  LRGs normalized to $z=0.2$ taken from \citet{wake06} (blue line). In
  order to compare the absolute magnitudes of the galaxies in our
  QSO-galaxy pairs sample, we overplot a histogram (light gray) line for
  all the galaxies selected.  The normalization of the histogram is
  arbitrary, but shows that the galaxies in our sample have similar
  luminosities to the LRG samples. The distribution of absolute
  magnitudes of galaxies that can be associated with Mg~II absorption are plotted in
  dark gray. The top x-axis shows the ratio of $L/L^*$ assuming $M_r^* +
  5\log h =  -20.4$ \citep{blanton03} and no evolution in the LRG LF between
  $z=0.1-0.2$. }
\end{figure}

We can compare the luminosities of the galaxies in the sample with
those of the LRG population. By adopting the same LRG SEDs discussed
above, we can derive suitable $K$-corrections to apply to our galaxies
and thereby estimate accurate absolute magnitudes $M$. These
corrections are necessary, since the values needed to correct, for
example, $M_r$ derived from $r$-band observations, and the true
rest-frame $M_r$, are $K_r \sim 0.9-1.4$ mags for $z=0.46-0.6$. The luminosity
function (LF) for LRGs at intermediate redshift has been derived by
\citet{wake06} and is reproduced in Figure~\ref{fig_lf}. The blue
points represent the volume density of LRGs at $0.5 < z < 0.6$ taken
from the 2SLAQ, corrected for passive evolution as well as including
$K$-corrections. In this case, the corrections to the absolute
magnitudes $M_{0.2r}$ are made to measure the luminosity at
$z=0.2$. Making the same correction, but not including an evolutionary
correction (which is much smaller than $K_r$), we measure the
distribution of $M_{0.2r}$ for the galaxies in our sample and plot
them as a histogram in Figure~\ref{fig_lf}. The actual normalization
of the histogram is not relevant here: the figure simply demonstrates that the
galaxies have largely the same distribution of luminosities as the LRG
population.

For the final sample of galaxies, we searched for existing (DR7) SDSS
spectra of the galaxies, in order to confirm both their photometric
redshifts and the shape of their SEDs. Only two spectra were
available, for two galaxies at distances of more than 144~kpc from
their background QSO sightlines, neither of which were associated with
any Mg~II absorption lines. The S/N of both spectra were very poor,
and insufficient to provide accurate redshifts. Given the limit of the
SDSS {\tt MAIN} galaxy sample, $r=17.8$, and the extended LRG sample,
$r=19.2$ \citep{eisenstein01}, we would expect very few of our
galaxies to have been observed spectroscopically: our sample consists
of only [0,0.3]\% of galaxies brighter than $r=$[17.8,19.2].

\section{Results \label{sectResults} }\label{sectResults}

Having applied all the selection criteria discussed in
\S\ref{sect_sample}, the final sample consisted of 593 QSO-galaxy
pairs.  There are 26 galaxies that have another galaxy with a
photo-$z$ lying closer to the QSO sightline, or, 23 galaxies that have
another galaxy with a photo-$z$ lying closer to the QSO sightline {\it
  and} for which the background QSO spectrum has $W_{\rm{lim}} <
0.6$~\AA .  For the remainder of this section, we exclude galaxies
further away than the  identified galaxy (with a photo-$z$),
but we reconsider if their exclusion is warranted in
\S\ref{sect_interlope}.  Of the 567 remaining galaxies, [175,42] have
impact parameters of $\rho\leq\:$[100,50]~kpc. At these separations,
we found Mg~II absorption within $\Delta z = \pm 0.1$ of the galaxies'
\ccz\ in [41,9] cases.  Representative spectra have already been shown
in Figure~\ref{fig_spectra}.  Of the 89 detections of Mg~II absorption
(at all radii), 11 spectra showed 2 or 3 Mg~II systems within the
search window.  This number is too small to infer anything about the
clustering of Mg~II systems in this sample, and more comprehensive
discussions on the cross-correlation between Mg~II systems and LRGs
have been given by \citet{bouche06} and \citet{tinker08}. In this
paper, we select the system with the redshift closest to the photo-$z$
of the LRG as the one ``associated'' with the galaxy or its
environment.

\begin{figure}
\vspace*{-1cm}\hspace*{-1cm}\includegraphics[width=11cm]{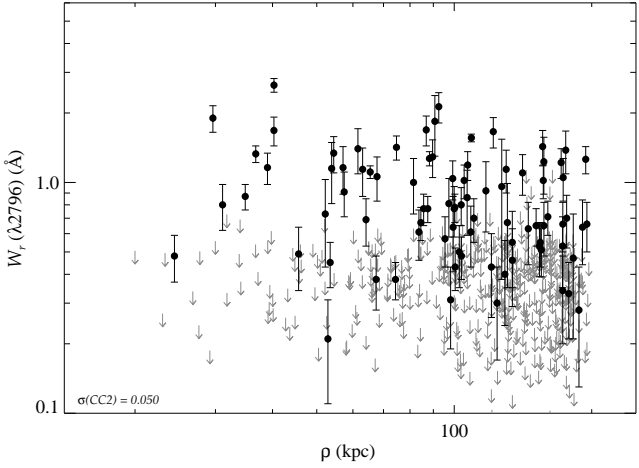}
\caption{\label{fig_ew1} Mg~II~$\lambda 2796$ eest equivalent width $W_r$, or
  $2\sigma$ limits, plotted against QSO-galaxy impact parameter $\rho$.} 
\end{figure}

The REWs, $W_r$, for detected Mg~II~$\lambda2796$ lines, or $2\sigma$
REW limits $W_{\rm{lim}}$ when no lines were detected, are plotted against galaxy
impact parameter $\rho$ in Figure~\ref{fig_ew1}. 
For each $\rho$, there is an uncertainty $\sigma(\rho)$ caused by not
exactly knowing the redshift of the galaxy. This error is small
though, since $|\sigma(\rho)| = 2 \rho \sigma(z)/(1+z)^2$, which is
only 7\% of $\rho$ for $\sigma(z) = 0.05$ at $z=0.5$. These errors are
omitted in Figure~\ref{fig_ew1}.
 The figure shows that
the EW limit of the survey is $\la 0.7$~\AA , as
expected from selecting QSO spectra with a minimum S/N ratio. That is,
we were able to always detect any Mg~II lines with $W_r > 0.7$~\AA
. In many cases of course, the spectra were sensitive enough to detect lines
much weaker than this, down to $W_r \ga 0.2$~\AA . Also as expected,
galaxy impact parameters extended out to $\approx 200$~kpc, the limit
imposed by selecting galaxies within 30 arcsec of a QSO
sightline.  Mg~II lines are detected up to REWs of $W_r \simeq 2$~\AA ,
but there is no correlation of $W_r$ with $\rho$. In particular, there
are many sightlines that show {\it no} Mg~II to REW limits of
$0.2-0.7$~\AA . Figure~\ref{fig_colors} demonstrates that Mg~II detections are
not a function of galaxy color, since detections are found for all
values of $g-r$ and $r-i$ colors in the sample.


\begin{figure}
\vspace*{-1cm}\hspace*{-0.9cm}\includegraphics[width=10.3cm]{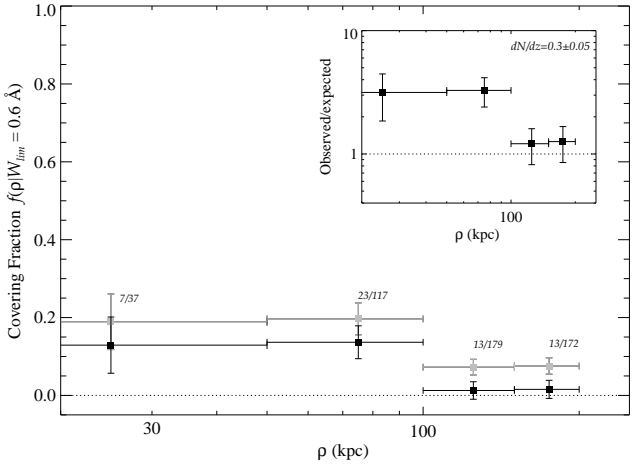}
\caption{\label{fig_cf} Plot of the covering fraction $f(\rho )$
  for galaxies in the sample, plotted against QSO-galaxy impact
  parameter $\rho$. The gray squares represent $f(\rho )$ without any
  corrections for the chance detection of an Mg~II system along a
  sightline; the fraction of detections to total number of pairs in
  each bin [which defines $f(\rho )$] is written next to each
  point. The black squares represent $f(\rho )$ after subtracting the
  predicted number of chance detections, assuming $dN/dz = 0.3$ at $z
  \simeq 0.5$. Horizontal error bars show the adopted bin sizes. If
  some of the absorption detected does not actually arise from the
  probed LRGs, but, e.g., from companion galaxies, then the points
  drawn here (and in Fig.~\ref{fig_all_lims}) represent only upper limits to
  $f(\rho$) for LRGs (see \S\ref{sect_interlope}). {\it Inset:} The
  ratio of the observed number of Mg~II systems to the expected
  number, again assuming $dN/dz = 0.3$.}
\end{figure}

The covering fraction of galaxies, $f(\rho )$, is defined simply as
the number of galaxies with detected Mg~II at some impact parameter,
compared to the total number of galaxies. To calculate $f(\rho )$, it
is necessary to use an EW limited sample of sightlines.
To that end, we selected only sightlines with S/Ns high enough
to reach REW limits $\leq
0.6$~\AA , irrespective of whether Mg~II was detected or not. This
meant that the selected spectra could detect lines as strong as, or
stronger than, 0.6~\AA . Since the highest S/N spectra in such a
sample could also
detect lines weaker than 0.6~\AA, we only counted a detection if it
too had a REW $\geq 0.6$~\AA. Hence the covering fractions include
sightlines where we {\it could} have detected lines with REW $\geq
0.6$~\AA , and where $W_r \geq 0.6$~\AA\ when a line was detected.
The
results are shown in Figure~\ref{fig_cf}, where values for $f(\rho )$
are plotted as gray squares. Error bars are taken to be simply
Poissonian errors on the number of detections. The covering fraction
is consistently low, starting at 0.2 between $\rho = 24-50$~kpc, and
declining to 0.1 between $\rho = 150-200$~kpc. 

Although we detect Mg~II within $\pm 2\sigma(z)$ of \ccz\ towards
some galaxies, in order to properly measure $f(\rho )$, we must also
consider how many of the Mg~II detections arise simply from
chance. The number of Mg~II absorbers per unit redshift, $dN/dz$, is
given by \citet{nestor05}, and at
$z=0.5$ is $\simeq 0.3$ for $W_r > 0.6$~\AA . From their
  work, we
estimate an error in $dN/dz \approx0.05$, and include this in
computing the error bars in  Figure~\ref{fig_cf}. Over a
given redshift interval, we can therefore calculate the number of
Mg~II absorbers expected to arise by chance and subtract these 
from the number of detections. The resulting values are shown as black
squares in Figure~\ref{fig_cf}. The ratio of the observed number of
systems to the expected number is also shown inset to
Figure~\ref{fig_cf}. The figure shows that $f(\rho )$ is small but
non-zero at $\rho < 100$~kpc, but that beyond this, $f(\rho )$ is
consistent with zero. This implies that any detection of Mg~II beyond
$\rho > 100$~kpc is probably just from chance, and that the galaxies
selected in this sample have no associated Mg~II absorption beyond
100~kpc.

\begin{figure}
\vspace*{-1cm}\hspace*{-0.8cm}\includegraphics[width=10.3cm]{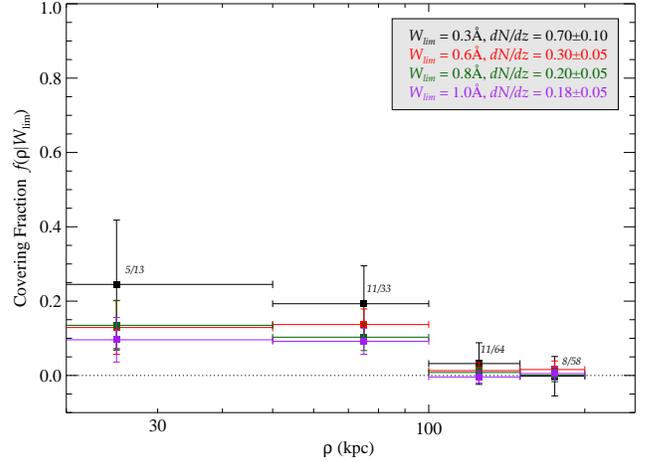}
\caption{\label{fig_all_lims} Plot of the covering fraction
  $f(\rho)$ for different values of  $W_{\rm{lim}}$, after correcting
  for the predicted number of Mg~II absorbers along the QSO
  sightlines. For the $W_{\rm{lim}} = 0.3$~\AA\ sample alone (shown
  with black points),
  the number of detections to the total number of sightlines selected
  is indicated as a fraction next to each point. The legend in the top
right corner matches the color of the points to the $W_{\rm{lim}}$
adopted in calculating  $f(\rho)$, and shows the values of $dN/dz$
used to correct the observed fraction of detections to non-detections.}
\end{figure}

We initially selected an REW limit of 0.6~\AA\ for two reasons: first,
because the limit lies close to the median of the REW distribution in our
sample, it would allow us to include a significant number
of spectra; and second, because a value of $dN/dz$ for that limit was
available in the literature.   In their search for Mg~II from a sample
of galaxies defined to be LRGs,
\citet{gauthier10} calculated a covering fraction for lines with a REW
of 0.5~\AA , a value that is similar to ours. They found $f(\rho ) \sim
20-40$\% for $\rho \la 100$~kpc, and a largely constant (with some
deviations) $f(\rho )$ of $\sim 20$\% between $\rho \simeq 100 -
400$~kpc. This result is clearly different from our measurement of a
sharp decline of $f(\rho )$ to approximately zero beyond 
100~kpc.

\citet{gauthier10} also calculated $f(\rho )$ for much stronger lines,
those with REW limits $> 1.0$~\AA , and found a factor of $\approx 2$
smaller $f(\rho )$ than for their analysis using REWs limits of 0.5~\AA
. We can see whether any differences in  $f(\rho )$ exist for
different REW limits in our sample. 
Since we are primarily
interested in $f(\rho)$ after correcting for Mg~II lines that might
arise by chance in our $\Delta z$ window, we need to know
$dN/dz$ at these REW limits. For $W_{\rm{lim}}$ greater than 0.6~\AA ,
we can use values of $dN/dz$ computed by \citet{nestor05} and
\citet{lundgren09} of $dN/dz = 0.2$ and $0.18$ for $W_{\rm{lim}} =
0.8$ and 1.0~\AA , respectively. Again, $f(\rho)$ depends weakly, but
non-negligibly, on the error in $dN/dz$, so we have estimated an
uncertainty of $\approx 0.05$ in $dN/dz$. The results are shown in
Figure~\ref{fig_all_lims}. For $W_{\rm{lim}} = 0.6 - 1.0$~\AA ,
$f(\rho)$ is largely the same, reaching between $10-15$\% for $\rho <
100$~kpc, then falling sharply at larger $\rho$.

We can also estimate $f(\rho)$ for $W_{\rm{lim}} < 0.6$~\AA
. Nestor~\etal\ measure $dN/dz \simeq 0.7$ for $W_{\rm{lim}} =
0.3$~\AA , although the error on this value appears to be larger than for the
other values of $dN/dz$ --- we estimate an error of $\approx 0.1$.
Figure~\ref{fig_all_lims} shows that at this REW limit, $f(\rho)$ may
be perhaps as much as a factor of two larger at $\rho < 100$~kpc,
although it too falls to zero quickly at $\rho > 100$~kpc.
Unfortunately, we have far fewer spectra in our sample with a S/N that
satisfies the limit of $W_{\rm{lim}} = 0.3$~\AA . The errors are
correspondingly large, and we cannot be sure that the larger $f(\rho)$
values arise merely as a consequence of poor statistics. To emphasize
this point, we label in Figure~\ref{fig_all_lims} the actual fraction
of Mg~II detections to the total number sightlines in each bin (as
also shown in Fig.~\ref{fig_cf}) for this particular $W_{\rm{lim}}$. So for $0 <
\rho < 50$~kpc, there are only 13 sightlines available, and we detect
Mg~II in five of them.  In the interval $50 < \rho
< 100$~kpc, we find Mg~II in 11 of 33 sightlines. In both cases, the errors
are large and compatible with the same values of $f(\rho)$ found
for higher $W_{\rm{lim}}$.

Searching for Mg~II lines near a REW limit of 0.3~\AA\ introduces
potential biases.  At high REW limits, 
the Mg~II doublet ratio, DR$ = (W_r\: \lambda
2796) / (W_r\: \lambda 2803$), of detected systems is expected to remain
low, close to unity, since both lines are likely to be (close to
being) saturated. This means that
if a $\lambda 2796$ line is detected at some significance (in our
case, at a $2\sigma$ level), then the $\lambda 2803$ will often be
detected at a similar significance. At a low $W_{\rm{lim}}$ of 0.3~\AA
, however, the Mg~II DR is likely to be higher (anywhere between 1.0
and 2.0), and a $\lambda 2796$ line with a strength close to
$W_{\rm{lim}}$ may be detected, but the $\lambda 2803$ line may
not be\footnote{Mg~II DRs are actually difficult to accurately measure in
  these SDSS data, since the fractional errors on both of the $W_r$ lines
  are relatively high, leading to large errors in DR. Although 
  there is a clear trend for systems with $W_r >0.6$~\AA\ to have DRs
  closer to 1.0, the number of detections for $W_r >0.3$~\AA\ is too
  low to provide a good comparison of DRs between the two samples.}. 
In our selection process, we required that both members of
the doublet be detected at a $2\sigma$ significance level, in order to
reject lines from other species at different redshifts. This selection
process could, in principle, lead us to under-estimate the number of
weak systems at low REW limits with $W_r \sim W_{\rm{lim}}$.
This would then increase the difference
between $f(\rho$) for $W_{\rm{lim}} = 0.3$~\AA\ and for higher
$W_{\rm{lim}}$ searches in Figure~\ref{fig_all_lims}.

A comparison of $f(\rho)$ for different REW limits with those
found by \citet{gauthier10} is complicated by the fact that the REW limits
may be defined slightly differently for the different analyses; for
example, the number of pixels over which a line is measured may be
different for each analysis. Although there is a suggestion in our
data that $f(\rho)$ changes by a factor of 2 as $W_{\rm{lim}}$ changes
from 0.3 to 0.6~\AA , there is no change in $f(\rho)$ for any increase
in $W_{\rm{lim}}$
when $W_{\rm{lim}} > 0.6$~\AA . Gauthier et al also see an increase in
$f(\rho)$ by a factor of 2, but for different REW limits, 0.5
and 1.0~\AA . It is possible that we are seeing the same change, and
that the values of $W_{\rm{lim}}$ are different simply because of
the differences in the way REWs are measured. We find it hard to
understand, however, why the differences in $W_{\rm{lim}}$ should be
so large, particular since both analyses use SDSS error arrays and
2$\sigma$ limits. We also continue to see the sharp decline in
$f(\rho)$ beyond 200~kpc, which Gauthier et al do not see for both
$W_{\rm{lim}} = 0.5$ and 1.0~\AA .

\begin{figure}
\vspace*{-1.5cm}\hspace*{-1.75cm}\includegraphics[width=11.5cm]{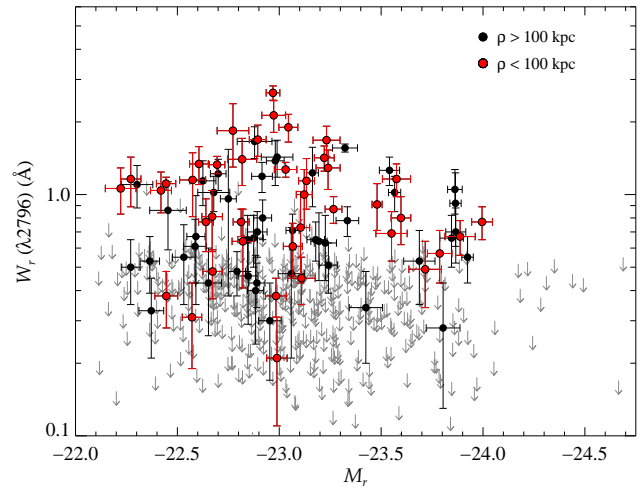}
\caption{\label{fig_absmag} Plot of Mg~II~$\lambda 2796$ against the
  absolute magnitude $M_r$ of the galaxies in our sample. Galaxies at impact
  parameters of $\rho < 100$~kpc are plotted in red, while the
  remainder are shown in black.  $K$-corrections were calculated for
  each galaxy using the LRG SED template discussed in
  \S\ref{sect_gals} (assuming no evolution) and \ccz .}
\end{figure}

Figure~\ref{fig_absmag} plots the REW of detected Mg~II absorption
lines, or limits, against the absolute magnitude of the galaxies. In
this case, the absolute magnitude is $K$-corrected to a redshift of
zero (as opposed to a rest frame of $z=0.2$ discussed in
\S\ref{sect_gals}). The figure shows no clear relationship between the
luminosity of a galaxy and the REW of Mg~II when detected. The figure
also shows the Mg~II detections segregated by $\rho$, with galaxies
associated with Mg~II detections at $\rho < 100$~kpc shown as red
circles, and galaxies with associated absorption at $\rho > 100$~kpc
shown as black circles. Again, there is no obvious difference between
the two groups.

\begin{figure}
\vspace*{-1cm}\hspace*{-1cm}\includegraphics[width=10cm]{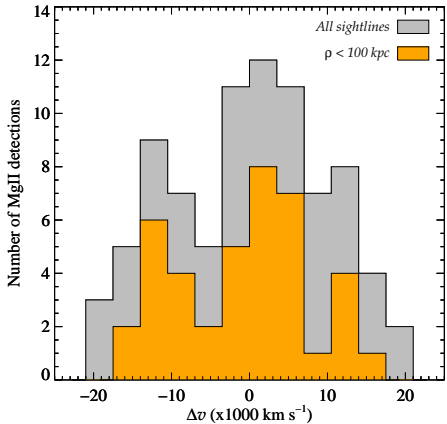}
\caption{\label{fig_delv} Histogram of the difference between the
  photometric redshift of the galaxy, \ccz, and the absorption
  redshift $z_{\rm{abs}}$ of the Mg~II system, when detected. The gray
histogram shows the distribution for all sightlines where Mg~II is
detected, whereas the orange histogram shows the distribution for
those pairs with $\rho < 100$~kpc.}
\end{figure}

The distribution of the difference between the photometric redshift of
the galaxy, \ccz, and the absorption redshift $z_{\rm{abs}}$ of the
Mg~II system, $\Delta v = c \Delta z / (1+$\ccz), is shown in
Figure~\ref{fig_delv}.  We do not expect this distribution to have a
high enough velocity resolution to distinguish between absorption from
individual LRG halos, or absorption from lower mass halos that cluster
around the larger mass LRG halos (for example). Indeed, although there
is some suggestion that the values of $\Delta v$ tend to cluster
around zero \kms, any peak to the distribution is poorly defined, as
we would expect given the photo-$z$ error associated with \ccz.  A
plot of $\Delta v$ against $\rho$ (not shown here) also shows no
convincing evidence that $\Delta v$ increases with $\rho$.

\section{Discussion}\label{sect_discussion}

In this paper, we have constructed a sample of QSO-galaxy pairs from
SDSS DR6 based on galaxies having photometric redshifts of between
$z=0.46-0.6$, and with small photometric-redshift errors of $\sim
0.05$. These photo-$z$s are taken directly from the SDSS database, and
are based on work described by \citet{oyaizu08}. Due largely to
selecting galaxies with small photo-$z$ errors, as well as needing to
use galaxies with magnitudes near the detection limit of the SDSS
imaging data, we have selected a sample made up primarily of LRGs,
as defined by their colors and luminosities.  We have
searched for Mg~II absorption in the SDSS spectra of the background
QSOs over a window $\Delta z = \pm 0.1$ of the photo-$z$ of the
galaxies. Our most significant result is that the detection of a Mg~II
system by an LRG is rare, even at small impact parameters of $\rho <
100$~kpc. A simple calculation of the covering fraction as a function
of $\rho$ suggest that the probability of detecting strong ($W_r >
0.6$~\AA ) lines within $\approx 20-100$~kpc\footnote{This lower limit
  of 20~kpc is set by the SDSS's ability to separate QSOs and galaxies
  in the imaging data; 20~kpc at $z=0.5$ correspond to an angular
  separation of 3.3 arcsec. We implicitly assume that LRGs have not
  been probed at radii less than 20~kpc in this work.} of an LRG is
only $10-15$\%.

\subsection{The Role of Unseen Interlopers \label{sect_interlope}}

Interpreting the covering fraction $f(\rho)$ requires some care. The
most obvious question is whether or not the covering fractions shown
in Figures 6 and 7 represent only lower limits to the true values of
$f(\rho)$.  For a galaxy at any particular $\rho$, if Mg~II is not
detected, then the result is clear: the absence of absorption is
completely valid for the LRG, even if there exists another galaxy at a
similar redshift which lies closer to the QSO sightline. Under the
assumption that the photo-$z$ of the galaxy is correct, then the
absence of absorption from that galaxy is unambiguous.

However, when Mg~II {\it is} detected, our selection process does not
rule out the possibility that another galaxy lies closer to the
sightline and is actually responsible for the absorption. 
Conventionally,
surveys to find absorbing galaxies are conducted with the aim of
identifying all
galaxies in a field, to some stated magnitude limit and
impact parameter. In our survey, selecting galaxies
based solely on having small errors in their photo-$z$s does not take
into account the possibility that a galaxy exists closer to the
sightline, and may be causing the absorption.  Late-type sub-$L^*$
galaxies are not easily detected in SDSS imaging data at the redshifts
of the LRGs: assuming $M^*_r = -21.2$ \citep{blanton03} a late-type
$L^*$ with a flat continuum below a rest-frame wavelength of 4000~\AA\
(for which a negligible $K$-correction would be required) would have
an observed magnitude of $r\simeq 21$ at $z=0.5$, about 1 mag brighter
than the 
detection limit of the SDSS. A visual inspection of the fields around the
QSO-LRG pairs in the SDSS data shows that the LRGs are often
accompanied by very faint galaxies with intermediate colors, some of
which may even lie closer to the QSO than the LRG. Unfortunately,
these objects have either no photo-$z$s, or photo-$z$s with very large
errors, making it difficult to generalize about their association with
the identified LRGs. In some fields, the faintest objects are not
identified in the SDSS database, and even 
brighter galaxies are not
detected if they overlap the QSO itself. An example of this can be
seen in Figure~\ref{fig_pairs}; although the LRG
J112013.16+671116.2 lies only 38~kpc from the QSO J112012.11+671115.9,
a galaxy can be seen due south of the sightline which is overlapping
the QSO sightline. The galaxy is not cataloged by SDSS, and no
photometric information is available. Mg~II is detected at the
redshift of the LRG, but whether this, or the interloping
galaxy is responsible for the absorption, is not known. The galaxy may
be unrelated to the LRG, or could be interacting with it. Without more
precise redshift information for all the galaxies in the field, and
deeper, higher resolution images, an understanding of this absorption
system is far from complete. Finally, we note that the presence of
other faint galaxies close to a given QSO-LRG pair is not simply confined
to fields where Mg~II is detected. Fields where Mg~II is not detected
are sometimes equally complicated. 

If other, gas-rich galaxies closer to the QSO sightlines than the LRGs
are indeed responsible for detected Mg~II absorption, then our association
of absorption systems with the LRG would, obviously, be erroneous and
would provide incorrect information on the absorbing characteristics
of LRGs. In Figures~\ref{fig_cf} and \ref{fig_all_lims}, therefore, we
would need to remove such sightlines from the appropriate $\rho$
bin. As this would lead to a lower fraction of detected absorption to
total sightlines, we would infer that the observed $f(\rho)$ were only
upper limits to the true value of $f(\rho)$ for LRGs.

Our inability to identify potential interlopers is the necessary
trade-off needed to study large numbers of galaxies using their
photometric redshifts instead of completing expensive redshift surveys
of many individual fields. 
Instead of considering which galaxies are nearest to a
sightline, 
an alternative approach to analyzing our
data might be 
to simply estimate the probability of finding Mg~II
absorption around any LRG, regardless of its environment. The plots of
$f(\rho)$ in Figures~\ref{fig_cf} and \ref{fig_all_lims} effectively
measure this probability, if we add back all the LRGs that were
excluded because another LRG had been identified closer to the
sightline (\S\ref{sectResults}). In fact, leaving in these LRGs makes
practically no difference to $f(\rho)$; of the 23 LRGs that have an
LRG closer to a line of sight (and for which $W_{\rm{lim}} \leq
0.6$~\AA ) only five show associated Mg~II, and their impact
parameters are between $90-183$~kpc, a range of $\rho$ that is dominated by
non-detections of Mg~II. This small number of detections adds a
negligible amount to $f(\rho)$ over this range.

\subsection{Possible Reasons for the Lack of Mg~II Absorption \label{sect_why_noabs}}

As discussed in \S1, LRGs are believed to inhabit some
of the highest
over-dense regions in the universe. In Halo Occupation Distribution
(HOD) models, they reside in massive halos with DM masses of
$10^{13-14}$~\msol , and are usually the central galaxies of rich
groups and clusters \citep{zheng09}. These high masses have been
confirmed through weak lensing measurements by
\citet{mandelbaum06}. \citet{ho09} have shown that X-ray luminous
clusters with virial masses of $1-5 \times 10^{14}$~\msol\ at $z\sim
0.2-0.6$ contain LRGs, and that the brightest LRGs lie at the centers
of the clusters. Other LRGs within a cluster, however, need not be
found near a cluster's center. 

A simple explanation for the lack of Mg~II from galaxies in our sample
is that the gas in the halos of LRGs is too hot to sustain clouds cool
enough to produce Mg~II absorption.  High mass halos are traditionally
expected to gain gas through ``hot mode'' accretion, whereby gas from
the intergalactic medium (IGM) is accreted into DM halos and is shock
heated to its virial temperature, $\sim 10^{7-8}$~K for the most
massive cluster halos
\citep[e.g.][]{rees77,white78,white91}. Additional feedback
mechanisms, such as supernovae driven winds, or AGN accretion, can
help maintain high gas temperatures around galaxies in massive halos
\citep[see, e.g. the review by][]{baugh06}.  For gas that is
collisionally ionized and in thermal equilibrium, most magnesium is in
the form of Mg~III at $10^5$K \citep{sutherland93}.  Lower mass halos
hosting $\sim L^*$ galaxies also have virial temperatures that are too
high for Mg~II to dominate the ion fraction \citep{fukugita06}, but
models in which cool photoionized gas can survive in the hot halo are
able to reproduce many of the characteristics of Mg~II (and other ion)
systems \citep[e.g.][]{mo96}.  Semi-analytical models and numerical
simulations of gas hydrodynamics in CDM constructs also find that cool
gas can exist in galaxy halos through ``cold mode'' accretion, which
occurs when gas [with, e.g., $T< 2.5\times 10^{5}$~K in
\citet{keres09}] is funneled from the IGM down filaments into the
halos. This process is expected to dominate in low mass ($<
10^{11-12}$\msol ) halos \citep{birnboim03, katz03, keres05,
  birnboim07, keres09}.

Several investigators have considered whether the masses of Mg~II
absorbing halos are consistent with those expected for the existence
of Mg~II absorbing clouds.  \citet{bouche06} cross-correlated Mg~II
absorption systems detected in SDSS spectra with a catalog of LRGs
(with photo-$z$s), from which they were able to measure a mean halo
mass $20-40$ times smaller than LRG halos. They found that the halo
mass was {\it anti-correlated} with Mg~II EW, and hence
that the absorbing clouds were not virialized in the halo.  This
result was unexpected, since more massive virialized halos, with
brighter galaxies, should have absorbing gas clouds with larger
velocity dispersions. Since the EW of a Mg~II line is,
to first order, proportional to the number of narrow components
comprising the line (at least for strong lines), then (assuming that
Mg~II absorption is directly related to clouds within a galactic
halo), $W_r$ should be proportional to the virial velocity of gas in
the halo, and hence to halo mass.  Bouch\'{e}~\etal\ estimated the
mass of halos in which strong ($W_r > 1$~\AA ) Mg~II was found to be
$\approx 10^{12}$~\msol , and determined that an anti-correlation
existed between halo mass $M_h$ and $W_r$.  \citet{lundgren09} refined
these values with additional SDSS data releases, and determined that
Mg~II absorbers with $W_r \geq 1.4$~\AA\ arose in halo
masses of $10^{11.3}$\msol, while absorbers with $0.8 \leq W_r \leq
1.4$~\AA\ arose in halos with $M_h \sim 10^{12.7}$\msol .

Our
results are in broad agreement with this anti-correlation of $W_r$ and
$M_h$, in the sense that we find few Mg~II systems in the high mass
halos of LRGs. Bouch\'{e}~\etal\ did not predict a covering factor for
a given halo mass, making it difficult for us to compare in detail our
results with their models. 
In interpreting their results, Bouch\'{e}~\etal\ favored the
explanation that galactic outflows within the halos
caused the majority of Mg~II systems, and not infalling IGM gas. In
keeping with this hypothesis, we would say that we see little Mg~II
absorption because few star-forming galaxies exist in high mass groups
and clusters. The LRGs in our sample are likely to be cluster
ellipticals, with little significant star formation or
outflowing cool gas. We return to this point below.

\citet{tinker08} and \citet{chen08} modeled Mg~II absorbing halos in
terms of the HOD formalism, and found
that in order to reproduce the anti-correlation of $W_r$ and $M_h$
seen by Bouch\'{e}~\etal , their models needed to incorporate a
transitional halo mass at which shock heated gas becomes more dominant
in higher mass halos, leading to less Mg~II absorption.
  In these models,
 the halos of Mg~II absorbing galaxies have
masses of $\sim 10^{11-13}$\msol ,
with a transitional
mass of $10^{11.6}$\msol , a result consistent with the
simulations of cold-mode accretion of gas into galaxy centers.  Again, in
the context of our survey, Mg~II is not detected near LRGs
because the galaxies reside in halos that are too massive to support cool
Mg~II clouds.

In detail, it is
less clear whether our results are compatible with Tinker \& Chen's
models, which were calculated for halo masses as high as
$10^{14}$\msol .  They found that ``a significant fraction of
cluster-sized halos contain some cold gas with a high covering
factor'', although they noted that the constraints on their models
were poor for halo masses $>10^{13}$\msol . If we take all
the LRGs in our sample to represent cluster-sized halos, our results
would tend to argue against these findings.  We would say
that our sample shows that a significant fraction of cluster-sized
halos contain cold gas with a low covering factor, or,
alternatively, only a few cluster-sized halos contain cold gas with a
high covering factor.

\subsection{Possible Origins for Observed Mg~II Absorption \label{sect_why_abs}}

Despite the fact that most LRGs do not cause Mg~II absorption, we
have shown that even after removing a contribution to the number of
detections that arise from chance along any sightline, there are still
a small number of Mg~II systems associated with the identified
galaxies. 
In this section 
we discuss possible origins for
these systems.

Could the few Mg~II systems we see around LRGs arise from the small
amounts of cool gas that remain in galactic halos?  Cold-mode
accretion is expected to be a dominate channel for galaxies at high
redshifts, or low-mass galaxies at lower redshifts. As indicated
above, the halo masses of LRGs are likely to be several dex above the
transition mass between the hot- and cold-mode regimes
($10^{11-12}$\msol ), so cold accretion is unlikely to be significant.
At $z=0$, numerical simulations suggest that the fraction of gas in
halos with masses of $10^{13}$\msol\ and temperatures
$<2.5\times10^{5}$~K is only a few percent, although whether cold
filamentary streams reaching into the halo cores at $z\sim 0.5$ are
able to survive is unclear \citep{keres09}.  The problem with this
speculation is that we do not actually know the mass distribution of
the LRGs we have sampled. Without a better measurement of the LRGs'
masses and environments, it is difficult to compare our results with
simulations and determine whether even small amounts of cold gas might
be associated with the LRGs we have studied.

A different explanation is that Mg~II absorption might arise in the
disks of interloping late-type galaxies which inhabit the outer
regions of the LRG groups and clusters \citep[e.g.][and refs.
therein]{hansen09}. Any difference in the redshifts between an LRG and
another cluster or group member will be much smaller than the error in
the photometric redshift of the LRG, and will not be detectable in
Figure~\ref{fig_delv}.  The morphologies of galaxies are known to
change from late- to early-type as the distance to the center of a
cluster decreases, and late-types may only need to present a small
cross-section to account for the low covering factor seen in
Figure~\ref{fig_cf}. Moreover, the transformations that take place in
late-type galaxies in the environment of rich groups and clusters
offer plausible mechanisms for increasing their Mg~II cross-section;
tidal interactions with other cluster galaxies, or stripping of
galactic ISMs by the intracluster medium, can re-distribute cool gas
over larger areas than those covered by single galaxy disks. The
problem with this scenario, however, is that we would expect the
surface density of late-type galaxies (as a function of the projected
distance from the center of a cluster) to be relatively constant over
Mpc scales\footnote{For example, the mean radius at which the galaxy
  density is defined to be 200 times the field density ($R_{200}$) is
  usually given to be $\simeq 1$~Mpc.}; there is no obvious reason why
the number of spiral galaxies would decline to zero at impact
parameters of 200~kpc from an LRG. Put another way, if late-type
galaxies at the outer regions of galaxy clusters were responsible
for detected Mg~II systems, we would expect the number of systems
``accidentally'' intercepted by QSO sightlines through a cluster to be
largely constant between 20 and 200 kpc of an LRG, possibly at a level
slightly higher than the background Mg~II systems discussed in
\S\ref{sectResults}. Since this is not what we observe, we consider
this scenario to be an unlikely explanation for the few Mg~II systems we do
detect.

A more likely explanation is that the Mg~II systems arise as a
consequence of wet mergers between LRGs and gas-rich satellites, with
the latter providing fuel for star formation in the former.  If
hierarchical merger models are correct, LRGs should certainly have the
most vigorous merger histories of galaxies in the universe. Again,
galaxy-galaxy interactions, or galaxy cannibalism, is a good way to
distribute tidal debris over areas of the sky larger than those of
individual galactic disks.  Absorption from tidally disrupted gas
  would also explain the large Mg~II EWs, since the
  kinematics of the interacting galaxies could spread Mg~II components
  over large velocity intervals \citep[e.g.][]{bowen_93j}.  More
importantly, however, if Mg~II systems are more closely associated
with star-forming galaxies, as discussed in \S1, where galactic
outflows push cool gas out of galactic disks, then perhaps LRGs that
have undergone recent bursts of star formation may be responsible for
strong Mg~II absorption lines. Again, outflows from star-forming
galaxy disks would explain the strong Mg~II EWs
\cite[e.g.][and refs.~therein]{chelouche10}.

The star formation history of LRGs has been studied by several
authors. \citet{eisenstein03} found that 10\% of the SDSS LRG
population have significant [O~II] emission, while \citet{roseboom06}
found that 12.5\% of $L>3L^*$ LRGs have spectra with some combination
of [O~II] emission lines and strong H$\delta$ absorption lines,
indicative of ongoing, or recent, star formation.  The similarity
between this fraction of LRGs, and the fraction we find associated
with Mg~II absorption, is striking. Excluding their [O~II]
emission-only subset of galaxies, $\simeq 4$\% of Roseboom~\etal 's
LRGs show good evidence for star formation; as noted by both
Eisenstein~\etal\ and Roseboom~\etal , the detection of emission lines
alone without corroborating absorption features is ambiguous, since
[O~II] can arise from galaxies with AGN activity (such as LINERS),
rather than from star formation.  The Mg~II absorption cross-section
of AGN has not been studied in detail, although QSOs at similar
redshifts to the LRGs in our sample are known to show high Mg~II
covering factors ($\sim 90$\%) out to $\sim 100$~kpc for lines with
$W_r > 0.5$~\AA\ \citep{bowen06,bowen11}.  Of course, QSO and AGN
activity may itself be triggered by galaxy mergers \citep[see,
e.g.][and refs therein]{hopkins08a}, a process which distributes gas
over a wide cross-section, triggers high star formation
\citep{mihos96}, and causes gas to be expelled from the galaxy through
supernovae driven winds \citep{springel05_bh}.  Hence, we may be seeing
Mg~II from the same combination of star-forming and (perhaps) AGN LRGs
as those selected by Roseboom~\etal\ and Eisenstein~\etal , leading to
the same fraction of Mg~II absorbers.  Roseboom~\etal\ were unable to
settle on the physical mechanisms which might produce star-forming/AGN
LRGs, but noted that their existence was consistent with
$\Lambda$CDM cosmologies that included mergers of gas-rich galaxies,
and cooling at the centers of clusters, much as already discussed
above.  Obviously, whether the LRGs in our sample show any
signs of the same star formation or AGN activity seen by
Eisenstein~\etal\ and Roseboom~\etal\ can be tested by simply
recording the spectra of the LRGs. 
\citet{gauthier10} have begun a
spectroscopic follow-up of their LRGs, but so far, only one galaxy
      associated with Mg~II absorption and at an impact parameter $<
      100$~kpc has been observed. Its SED shows no obvious signs
      of star-formation or AGN activity. 

Whether Mg~II absorption originates in tidal debris from merging
galaxies, or whether it arises in outflows from star formation {\it instigated} by
merging galaxies, in both cases the processes involve the interaction of galaxies. 
Unfortunately, at redshifts of $\sim 0.5$,
measuring the rate
of even {\it major} LRG mergers, where both interactors might be bright
enough to be visible in large-area sky surveys, has proved
difficult; the rate though, does appear to be low  \citep[][and refs
therein]{masjedi08,bundy09,depropris10}.  
The comoving volume density
of the most massive ($M> \sim 10^{11} M_\odot$) galaxies changes
little at $z\sim0.2-1$. The numbers are not increasing [at least to
within factors of $2-3$ --- see, e.g. \citet{conselice07}], even
though more would be expected to arise at lower-$z$ if they continued
to accrete galaxies through mergers. 

At much lower redshifts, $z\simeq 0.02-0.14$, where galaxy environment is
easier to map in shallow galaxy surveys, \citet{huang09} found that
5\% of bright (between $M_r$ of $-20$ and $-24$, so similar to the
LRGs) E and S0 galaxies showed signs of ongoing or recent star
formation, and an additional $10-13$\% showed LINER or Seyfert
activity. These numbers are similar to those measured specifically for
the LRGs by Roseboom~et~al.  Unfortunately, Huang \& Gu
excluded galaxies with disturbed morphologies, making it
hard to know how much star formation in early-type galaxies might be
associated with merging systems.  Interestingly, the environments of
their star-forming ellipticals appear to be quite mixed. Although 5
galaxies seem to be relatively isolated, 8 of the 13 listed by Huang
\& Gu have one or more neighbors with similar redshifts within
300~kpc
and/or have been associated with galaxy groups or clusters by other
authors. Hence, at low-$z$ it seems plausible that a small number of
bright, star-forming ellipticals can be found in similar dense
environments to those of LRGs at higher-$z$.

Can we measure any difference in the environment of those LRGs that we
associate absorption with and those that show no absorption? To
investigate this question, we collated all galaxies within 5 arcmin of
every QSO sightline in our survey from the SDSS DR7 {\tt GALAXY} database. At
$z=0.6$, 5 arcmins corresponds to a proper length of 2~Mpc, which
approximates cluster-like scales. We calculated the number of galaxies per
shell along each sightline, $n(r)$, where we counted galaxies in shells
of width 50 arcsec, and then compared both the average and median of
$n(r)$ for all sightlines  with and without Mg~II detections.
We took the error in these values to be the
measured standard deviation in $n(r)$. 
In one case we counted all galaxies,
regardless of their redshifts; in a second case, we used only red
galaxies that satisfied the color selection of \citet{cannon06}
discussed in \S\ref{sect_gals} and shown in
Figure~\ref{fig_colors}. In the latter case, these red galaxies are
more likely to be at the redshifts of the previously selected LRGs.

For both galaxy samples, unfortunately, we could not differentiate
between the average or median of $n(r)$ for sightlines with and
without Mg~II absorption. We found that the variance in the number of
galaxies per field dominates the distribution of $n(r)$ for both
absorbers and non-absorbers. In particular, having only 85 unique
QSO sightlines showing Mg~II detection provides too small a sample to show
any difference with the distribution of galaxies in the 482 QSO fields
that show no Mg~II. A larger 
study
 which matches all LRG environments
around all QSO sightlines, and not just QSOs with LRGs within 30
arcsec of a sightline (as used in this study), may be more fruitful,
but is beyond the scope of this paper.

Our search for Mg~II absorption lines from LRGs can only be used so
far in interpreting the origin of QSO absorption line systems.  The
lack of Mg~II from this population of galaxies suggests that the
majority of Mg~II systems have little to do
with massive red galaxies, nor probably, massive hot halos. The
implication is that if Mg~II systems are indeed associated with
galaxies (and are not simply metal-enriched intergalactic clouds) then
other types of galaxies and environments are primarily responsible for
Mg~II absorption. This is hardly a surprising conclusion, given the
identification of the strong links between Mg~II systems and
late-type, star-forming galaxies discussed in the
Introduction of this paper. Indeed, as our understanding of galaxy properties and
galaxy number counts has become more refined, and with a more precise
counting of Mg~II systems at the lowest redshifts, it has become easier
to account for the origin of {\it all} Mg~II systems by star forming
disk galaxies 
\citep[e.g.][and refs therein]{chelouche10}.  Whether the small
percentage of Mg~II absorbers found herein to be associated
with LRGs provides any deeper insights into the origin of Mg~II
systems, is not clear.

\acknowledgments

D.V.B is funded through LTSA NASA grant NNG05GE26G.  D.C acknowledges
support from a Marie Curie grant PIRG06-GA-2009-256434. We thank FG
and Chuli for their continuing support. This paper is dedicated to the
memory of FG.
Funding for the SDSS and SDSS-II has been provided by the Alfred
P. Sloan Foundation, the Participating Institutions, the National
Science Foundation, the U.S. Department of Energy, the National
Aeronautics and Space Administration, the Japanese Monbukagakusho, the
Max Planck Society, and the Higher Education Funding Council for
England. The SDSS Web Site is http://www.sdss.org/.

The SDSS is managed by the Astrophysical Research Consortium for the
Participating Institutions. The Participating Institutions are the
American Museum of Natural History, Astrophysical Institute Potsdam,
University of Basel, University of Cambridge, Case Western Reserve
University, University of Chicago, Drexel University, Fermilab, the
Institute for Advanced Study, the Japan Participation Group, Johns
Hopkins University, the Joint Institute for Nuclear Astrophysics, the
Kavli Institute for Particle Astrophysics and Cosmology, the Korean
Scientist Group, the Chinese Academy of Sciences (LAMOST), Los Alamos
National Laboratory, the Max-Planck-Institute for Astronomy (MPIA),
the Max-Planck-Institute for Astrophysics (MPA), New Mexico State
University, Ohio State University, University of Pittsburgh,
University of Portsmouth, Princeton University, the United States
Naval Observatory, and the University of Washington.


\bibliography{bib2}


\clearpage

\begin{sidewaystable}
\caption{Mg~II Absorption in QSO-galaxy pairs \label{tab_sample}}
\begin{tabular}{ccl lccccc crrr}
\hline
 & & & & & & & & & & & \multicolumn{2}{c}{$W_r$ (Mg~II)} \\
\cline{12-13}
 & & & & & & 
& $\theta$
& $\rho$                  
& 
& \push{$\Delta z$}              
& \push{$\lambda 2796$}
& \push{$\lambda 2803$} \\
QSO           
& Plate$-$MJD$-$fibre
& $z_{\rm{QSO}}$
& \push{Galaxy}
& \ccz
& $r$
& $M_r$
& ($''$)
& (kpc)                  
& $z_{\rm{abs}}$          
& (\kms)                 
& \push{(\AA )}
& \push{(\AA )} \\
(1)            
& (2)
& \push{(3)}
& \push{(4)}
& (5)
& (6)
& (7)
& (8)
& (9)
& (10)
& \push{(11)}
& \push{(12)}
& \push{(13)} \\
\hline
\input{table1}
\hline
\end{tabular}
\begin{minipage}[b]{21cm} 
\bigskip
{\it The complete version of this table can be found at
  http://www.astro.princeton.edu/$\sim$dvb/ST/lrgtable.pdf}. \\

  Entries
  are listed in order of increasing QSO-galaxy impact parameter
  (col.~9). In some
  cases, multiple galaxies satisfying our selection criteria
  were found around a single sightline. These are also listed, but are
  kept together with a single QSO entry, even though they may have
  much larger impact parameters.  
  The columns list the following parameters:
  Col.~(1): Background QSO
  designation. For brevity, the ``SDSS'' prefix has been dropped from
  this list, and the list in col.~(4); 
  col.~(2): spectroscopic fiber identification from DR6; 
  col.~(3): emission redshift of QSO;
  col.~(4): galaxy designation; 
  col.~(5): galaxy photometric redshift and error; 
  col.~(6): galaxy $r$-band model magnitude, de-reddened for Milky Way
  dust absorption;
  col.~(7): galaxy absolute magnitude, based on \ccz , after applying a $K$-correction
  to measure the $r$-band luminosity at a redshift $z=0$;
  col.~(8): angular separation between QSO sightline and galaxy;
  col.~(9): impact parameter between QSO sightline and galaxy;
  col.~(10): Mg~II absorption redshift, if doublet detected within
  $\pm0.1$ of \ccz;
  col.~(11): difference in redshift between $z_{\rm{abs}}$ and \ccz;
  cols.~(12,13): rest equivalent widths and errors of Mg~II lines, when
  detected, or $2\sigma$ upper limit, if absorption lines not detected.  
\end{minipage}
\end{sidewaystable}

\end{document}

%% file: table1.tex
J$010601.41-083556.8$ & $0659-52199-361$ & 1.384 & J$010601.17-083557.0$ & $0.476\pm0.028$ &  19.7 & $-23.1$ &  3.4 &  20.0 & $\ldots$ & $\ldots$ & $<\: 0.57$ & $<\: 0.57$ \\
J$123744.56+110658.2$ & $1233-52734-443$ & 0.946 & J$123744.60+110654.3$ & $0.469\pm0.046$ &  20.7 & $-22.0$ &  3.9 &  23.0 & $\ldots$ & $\ldots$ & $<\: 0.30$ & $<\: 0.30$ \\
J$122922.94+495650.5$ & $0971-52644-156$ & 1.600 & J$122922.71+495653.5$ & $0.508\pm0.035$ &  20.5 & $-22.6$ &  3.8 &  23.3 & $\ldots$ & $\ldots$ & $<\: 0.73$ & $<\: 0.74$ \\
J$115312.76+463553.5$ & $1446-53080-202$ & 1.123 & J$115312.88+463549.7$ & $0.495\pm0.039$ &  20.6 & $-22.4$ &  4.0 &  24.4 & 0.4417 & $ 0.0532$ & $ 0.47\pm0.13$ & $ 0.14\pm0.14$ \\
J$083426.64+291311.1$ & $1268-52933-493$ & 1.558 & J$083426.37+291313.4$ & $0.562\pm0.035$ &  20.4 & $-23.2$ &  4.1 &  26.8 & $\ldots$ & $\ldots$ & $<\: 0.48$ & $<\: 0.49$ \\
J$160624.23+434336.7$ & $1170-52756-399$ & 1.489 & J$160624.39+434340.8$ & $0.517\pm0.023$ &  19.5 & $-23.7$ &  4.4 &  27.6 & $\ldots$ & $\ldots$ & $<\: 0.42$ & $<\: 0.40$ \\
J$125929.93+560525.9$ & $1318-52781-203$ & 1.674 & J$125929.40+560525.2$ & $0.515\pm0.025$ &  20.1 & $-23.1$ &  4.5 &  27.7 & $\ldots$ & $\ldots$ & $<\: 0.64$ & $<\: 0.62$ \\
J$082840.61+364819.0$ & $0827-52312-143$ & 1.376 & J$082840.92+364815.8$ & $0.462\pm0.042$ &  20.1 & $-22.6$ &  5.0 &  29.2 & $\ldots$ & $\ldots$ & $<\: 0.23$ & $<\: 0.24$ \\
J$102054.61+501007.3$ & $1008-52707-251$ & 1.438 & J$102054.55+501012.2$ & $0.488\pm0.038$ &  20.1 & $-22.8$ &  4.9 &  29.6 & 0.4350 & $ 0.0531$ & $ 1.91\pm0.25$ & $ 1.25\pm0.26$ \\
J$162921.31+160356.8$ & $2207-53558-288$ & 1.010 & J$162921.50+160352.6$ & $0.479\pm0.035$ &  19.8 & $-23.0$ &  5.0 &  30.0 & $\ldots$ & $\ldots$ & $<\: 0.63$ & $<\: 0.63$ \\
J$143248.46+613817.4$ & $0607-52368-520$ & 2.017 & J$143248.01+613820.9$ & $0.554\pm0.029$ &  20.2 & $-23.3$ &  4.8 &  31.1 & 0.6155 & $-0.0618$ & $ 0.69\pm0.28$ & $ 0.81\pm0.28$ \\
J$145307.68+354200.1$ & $1384-53121-354$ & 1.752 & J$145307.96+354156.4$ & $0.534\pm0.037$ &  20.8 & $-22.6$ &  5.0 &  31.8 & $\ldots$ & $\ldots$ & $<\: 0.65$ & $<\: 0.63$ \\
J$142329.45+020443.0$ & $0534-51997-514$ & 0.961 & J$142329.36+020437.9$ & $0.491\pm0.042$ &  20.5 & $-22.4$ &  5.3 &  31.8 & $\ldots$ & $\ldots$ & $<\: 0.40$ & $<\: 0.41$ \\
J$140513.91+135219.2$ & $1704-53178-472$ & 1.085 & J$140513.66+135215.0$ & $0.487\pm0.036$ &  20.3 & $-22.6$ &  5.6 &  33.8 & $\ldots$ & $\ldots$ & $<\: 0.74$ & $<\: 0.71$ \\
$\ldots$ & $\ldots$ & $\ldots$ & J$140515.71+135221.4$ & $0.473\pm0.039$ &  20.7 & $-22.1$ & 26.4 & 156.4 & $\ldots$ & $\ldots$ & $<\: 0.75$ & $<\: 0.78$ \\
$\ldots$ & $\ldots$ & $\ldots$ & J$140515.43+135235.5$ & $0.491\pm0.024$ &  19.7 & $-23.2$ & 27.4 & 165.9 & $\ldots$ & $\ldots$ & $<\: 0.67$ & $<\: 0.72$ \\
$\ldots$ & $\ldots$ & $\ldots$ & J$140514.23+135150.6$ & $0.570\pm0.026$ &  20.0 & $-23.7$ & 29.0 & 189.2 & $\ldots$ & $\ldots$ & $<\: 0.53$ & $<\: 0.53$ \\
J$130351.25+141650.4$ & $1771-53498-064$ & 1.185 & J$130351.54+141646.8$ & $0.497\pm0.043$ &  20.3 & $-22.7$ &  5.6 &  34.0 & $\ldots$ & $\ldots$ & $<\: 0.60$ & $<\: 0.59$ \\
$\ldots$ & $\ldots$ & $\ldots$ & J$130351.41+141718.1$ & $0.464\pm0.040$ &  20.4 & $-22.3$ & 27.8 & 162.9 & $\ldots$ & $\ldots$ & $<\: 0.79$ & $<\: 0.81$ \\
J$091040.04+535427.9$ & $0553-51999-509$ & 1.096 & J$091040.59+535430.5$ & $0.511\pm0.031$ &  20.5 & $-22.6$ &  5.5 &  34.1 & $\ldots$ & $\ldots$ & $<\: 0.62$ & $<\: 0.61$ \\
J$122522.20+325630.8$ & $2015-53819-247$ & 1.119 & J$122521.79+325632.9$ & $0.530\pm0.035$ &  20.3 & $-23.0$ &  5.5 &  34.9 & 0.5246 & $ 0.0053$ & $ 0.72\pm0.18$ & $ 1.06\pm0.17$ \\
J$132332.32+560339.7$ & $1320-52759-180$ & 1.037 & J$132333.06+560340.2$ & $0.472\pm0.025$ &  19.7 & $-23.0$ &  6.2 &  36.7 & $\ldots$ & $\ldots$ & $<\: 0.71$ & $<\: 0.75$ \\